\documentstyle[11pt,aaspp4,flushrt,tighten]{article}
\singlespace

\font\gkvec=cmmib10                         
\def\bbeta{\hbox{{\gkvec\char12}}}         
\def\ni{{\noindent}}
\def\etal{{et al.\ }}
\def\spose#1{\hbox to 0pt{#1\hss}}
\def\lta{\mathrel{\spose{\lower 3pt\hbox{$\mathchar"218$}}
     \raise 2.0pt\hbox{$\mathchar"13C$}}}
\def\gta{\mathrel{\spose{\lower 3pt\hbox{$\mathchar"218$}}
     \raise 2.0pt\hbox{$\mathchar"13E$}}}
\def\np{n_{e^+}}

\def\gamav{\langle\gamma_e\rangle}
\def\bk{{\bf k}}
\def\bkone{{\bf k}_s}
\def\tiln{n'}
\def\wiell{\widetilde\ell}

\begin{document}
\title{Relativistic Winds from Compact Gamma-ray Sources:\break I. 
Radiative Acceleration in the Klein-Nishina Regime}

\author{Piero Madau\altaffilmark{1,2} and Christopher 
Thompson\altaffilmark{3}}

\altaffiltext{1}{Institute of Astronomy, Madingley Road, 
Cambridge CB3 0HA, UK.}
\altaffiltext{2}{Institute for Theoretical Physics, University of California, 
Santa Barbara, CA 93106--4030.}
\altaffiltext{3}{Department of Physics and Astronomy, University of North 
Carolina, Chapel Hill,
NC 27599.}

\begin{abstract}

\ni We consider the radiative acceleration to relativistic 
bulk velocities of a cold, optically thin plasma which is exposed to 
an external source of $\gamma$-rays. The flow is driven by radiative 
momentum input to the gas, the accelerating force being due to Compton 
scattering in the relativistic Klein-Nishina limit.   
The bulk Lorentz factor of the plasma, $\Gamma$, derived as a function of
distance from the radiating source, is compared with the corresponding result
in the Thomson limit. Depending on the geometry and spectrum of the
radiation field, we find that particles are
accelerated to the asymptotic Lorentz factor  at infinity
much more rapidly in the relativistic regime; and the radiation drag is
reduced as blueshifted, aberrated photons experience a decreased relativistic
cross section and scatter preferentially in the forward direction. 
The random energy imparted to the plasma by $\gamma$-rays can be converted into
bulk motion if the hot particles execute many Larmor orbits before 
cooling. This `Compton afterburn' may be a supplementary source of momentum 
if energetic leptons are injected by pair creation, but can be  
neglected in the case of pure Klein-Nishina scattering.
Compton drag by side-scattered radiation is shown to be more
important in limiting the bulk Lorentz factor than the finite
inertia of the accelerating medium.  The processes
discussed here may be relevant to a variety of astrophysical situations where
luminous compact sources of hard X- and $\gamma$-ray photons are observed,
including active galactic nuclei, galactic black hole candidates, and 
gamma-ray bursts. 

\end{abstract}


\section{Introduction}

The ejection of particles by radiation pressure has been 
considered many times as a possible mechanism for producing relativistic
outflows from very luminous radiation sources, such as active galactic nuclei
(AGNs) or galactic compact objects. In the case of an optically thin plasma, 
the particle dynamics is dominated by the intense photon field and a flow
results from the work done by the radiation force on the material.

In most previous studies it was assumed that the particles scatter
photons with a cross section independent of frequency and front-to-back
symmetric along the incident photon direction, such as true in the Thomson
limit (Noerdlinger 1974; O'Dell 1981; Phinney 1982; Kovner 1984). 
Direct momentum input to the fluid at rest results when outward-directed
photons are removed from the anisotropic radiation field, and then scattered 
away. No net momentum is carried off by the scattered photons and the material
is accelerated radially. A particle moving outward near the source, however,
encounters many photons nearly at right angles to its motion, more so as
$v\rightarrow c$ because of aberration effects, and suffers collisional 
drag by them. Only mildly relativistic flows can be produced by an 
Eddington-limited extended source, the radiation drag actually preventing 
acceleration to very high particle energies (Noerdlinger 1974).
In the context of AGNs, the role of Compton drag in the
dynamics of relativistic jets has been the subject of many studies 
(e.g. Abramowicz \& Piran 1980; Sikora \& Wilson 1981; Phinney 1987; 
Melia \& K\"{o}nigl 1989; Sikora \etal 1996). 

In the past few years, the {\it Compton Gamma-Ray Observatory} mission 
has revealed the 
existence of some classes of objects, such as the EGRET (Fichtel \etal 
1994), OSSE (McNaron-Brown \etal 1995), and COMPTEL (Bloemen \etal 1995)
blazars, together with some $\gamma$-ray pulsars (Ulmer 1994), where 
the hard X-ray and $\gamma$-ray fluxes are a significant fraction or 
completely dominate the overall radiation energy budget. This is similar 
to what sometimes observed by the SIGMA experiment onboard of {\it Granat} 
in the case of galactic black hole candidates (Mandrou \etal 1994). The large 
compactnesses inferred from these data naturally lead to theoretical models 
in which Compton scattering in the Klein-Nishina (KN) limit (as opposed to 
Thomson scattering), together with photon-photon pair production  may play a 
role 
in determining the thermal state and 
dynamics of the source. Such process may also be relevant for studies of the
effects of gamma-ray bursts on their gaseous environment (their 
``afterglows''). 

In this paper we study the dynamics of a {\it cold} plasma embedded in 
the ``hard" photon field of an external source. We adopt a test-particle 
approach, which is valid only insofar gas pressure gradients
can be neglected compared to the radiative force and the material is 
optically thin. We consider the model problem of a fully ionized plasma 
where gravity is overwhelmed by radiation pressure and the distribution of
particle momenta remains one-dimensional.  
We ignore stimulated scattering, photon absorption, and 
photon-electron pair production. 
Free-free absorption is unimportant compared to Thomson scattering if 
$\sigma_{ff}=8\times 10^{-47}n_e\,(h\nu/10^{-3 }m_ec^2)^{-7/2}\,{\rm cm^2}
<\sigma_T$ ($h\nu\gg kT$), where $n_e$ is the electron density.
Photo-electron pair production, $\gamma e^-\rightarrow 
e^+e^-e^-$ (threshold $h\nu'>4\,m_ec^2$, where $h\nu'$ is the photon energy
in the electron rest-frame) is a higher order 
process in the fine-structure constant, important only at large
energies.  (The pair production cross section 
is $\sim 30\%$ of the KN scattering cross section at $h\nu'=100\, m_ec^2$, e.g. 
Svensson 1982.) In an accompanying paper (Thompson \& Madau 1999, hereafter 
Paper II) we show, however, how  pair creation induced by
collisions between hard photons and soft side-scattered photon may 
increase the acceleration rate by decreasing the inertia per particle,
and by increasing the mean momentum deposited per scattering.  

The plan of this paper is as follows. In \S\,2 we discuss the basic
theory of radiative acceleration. In \S\,3  we present sample solutions for
extended and impulsive radiation sources in the Thomson regime.
Compton drag by side-scattered radiation is shown to be more effective
in limiting the bulk motion than is the finite inertia of the accelerating
medium.  The random energy imparted to the plasma by $\gamma$-rays can be 
converted into bulk motion if the hot particles execute many Larmor orbits 
before cooling. 
The equation  of motion in the KN limit is derived and numerically
integrated in \S\,4 for a monoenergetic spectrum, under the
assumption that the outflowing particles maintain a one-dimensional
distribution of momenta, i.e. that the plasma remains `cold' even in the
presence of recoil.
Blumenthal (1974) used the KN cross section to calculate
the mean force due to Compton scattering on electrons with arbitrary velocity.
An approximate solution to the problem of
the radiative deceleration of a relativistic jet in the KN regime has been
recently given by Luo \& Protheroe (1999).

\section{General formalism}

Consider an optically thin plasma exposed to a photon source.  The incident
and scattered photon momenta (in units of $m_ec$) are denoted by $x\bk$ and
$x_s\bkone$ in the (unprimed) lab frame.  Let $I_x(\bk)$ be the specific
intensity of the incident radiation in the direction $\bk$, 
$I(\bk)=\int I_x(\bk)dx$.  In the simplest case of a leptonic plasma
composed entirely of electrons and positrons, the bulk momentum is 
obtained directly from the normalized phase-space density $f(\bf p)$,
\begin{equation}
\langle \Gamma\bbeta\rangle = \int {{\bf p}\over m_ec}f({\bf p})d^3p.
\end{equation}
Here, $\Gamma=(1-\beta^2)^{-1/2}$ as usual.  This equation is easily
generalized to include a hadronic component.
The mean rate of momentum transfer per particle is then given by 
\begin{equation}
{\bf\cal  F}_{\rm rad} = \mu c{d\over dt}\langle \Gamma\bbeta\rangle =- 
{\sigma_T\over c} \int \left[{x_s{\bkone} -x{\bk} \over x}
(1-\bbeta\cdot \bk)I_x(\bk)dxd\sigma d\Omega\right]f({\bf p})d^3p. 
\label{eq:eqp}
\end{equation}
Here, $\mu$ is the mean mass per scattering charge,
the angular integral is over the solid angle subtended by the source, and
$d\sigma$ is the invariant differential cross section.  

When the number of scattering charges is dominated by pairs, we have
assumed implicitly a magnetic field strong enough 
to couple the pairs with the hadronic component of the plasma.
The minimum flux density needed to enforce that coupling is estimated
in \S \ref{seccompton}.   One must then describe the accelerating
particles as a fluid, with 
\begin{equation}
{d\Gamma\beta\over dt} \rightarrow {\partial\Gamma\beta\over\partial t}+
c\bbeta\cdot{\bf\nabla}(\Gamma{\beta}).
\label{eq:gamac}
\end{equation}
In a cold plasma composed of hydrogen (of density $n_p$) and $e^\pm$ pairs 
(of density $n_{e^+}$), 
\begin{equation}
\mu \approx m_e + m_p\left(1+2{n_{e^+}\over n_p}\right)^{-1}.\label{eq:mueq}
\end{equation}
This expression neglects the inertia of the neutralizing
electron component, but not of the pairs.

The Klein-Nishina cross section for scattering of unpolarized radiation
is most simply expressed in the (primed) electron rest-frame as
\begin{equation}
d\sigma ={3 \over 16\pi}\sigma_T {\left ({x_s'\over x'}\right )}^2 
\left[{x'\over x_s'}+{x_s'\over x'}-\sin^2\chi'\right]\sin \chi'd\chi'
d\phi ', \label{eq:eqkn}
\end{equation}
e.g. Rybicki \& Lightman (1979).  The energy of the scattered photon is
\begin{equation}
x_s'={x'\over 1+x'(1-\cos \chi')}; \label{eq:eqxp}
\end{equation}
in this notation, the scattering takes place through a polar angle
$\chi'$ and an azimuthal angle $\phi'$. 

Measuring the angles $\theta$ and $\theta_s$  of the 
incident and scattered photon with respect to the electron velocity $\bbeta$,
we can write the following kinematic relations 
\begin{eqnarray}
x_s & = & x_s'\Gamma (1+\beta \cos\theta_s') \nonumber \\ 
x & = & x'\Gamma (1+\beta \cos\theta') \nonumber \\
\cos\theta'& = & {(\cos\theta-\beta)\over (1-\beta \cos\theta)} \\
\cos\theta_s'& = & \cos\theta' \cos\chi'+\sin\theta' \sin\chi' \cos\phi'.
\nonumber 
\label{eq:eqkin} 
\end{eqnarray}
Since the scattering cross section is independent of the azimuth 
$\phi'$, we must integrate equation (\ref{eq:eqp}) over two angles, 
$\chi'$ and $\theta'$. The evaluation of the integral in equation 
(\ref{eq:eqp}) will be made by assuming that the radiation field is 
symmetric around the $\bf r$-axis, and the electrons (and positrons) will
maintain a one-dimensional distribution of momenta along $\bf r$, 
$f(p {\bf n})=f(p)\delta({\bf n}-\hat{\bf r})$.
Note that multiple scatterings can be safely neglected, as the singly scattered
photons are beamed into a cone of half-angle $\sim 1/\Gamma$ along the
electron direction of motion. Further scatterings are therefore suppressed
by the $(1-\bbeta\cdot {\bk})\sim (2\Gamma^2)^{-1}$ factor in the momentum
transfer equation. Single side-scattered photons will, however, be
absorbed via $\gamma + \gamma \rightarrow e^+e^-$ if the compactness
of the $\gamma$-ray source is high at $x\gta 1$. We defer a detailed
discussion of the effects of pair creation to Paper II.

\section{Thomson scattering}

To provide some physical insight and a framework to interpret our numerical
results in the relativistic limit, we shall first discuss the acceleration
of a test charge in the Thomson scattering approximation, $x' \ll 1$.
In this regime, $x'_s\approx x'$ by equation (\ref{eq:eqxp}).
Thus, in the electron rest-frame, photons of unchanged energies are
scattered into an angular distribution that is forward-backward symmetric
along the incident photon direction. 

A particle accelerated by an external photon source will, in general, follow
a curved trajectory.  The problem simplifies when the source is cylindrically
symmetric.  From equation (\ref{eq:eqp}), a parcel of matter moving along the
axis of symmetry with speed $\beta = c^{-1}dr/dt$ is accelerated at the rate
\begin{equation}
{d\langle\Gamma\beta\rangle\over dt}={\sigma_T \Gamma^2 \over
\mu c^2}\left[(1-\beta)^2 \int I(\theta)\cos \theta d\Omega -\beta \int
I(\theta)(1-\cos \theta)^2 d\Omega \right].\label{eq:eqT}
\end{equation}
The first term in brackets represents the collimated flux of redshifted photons
which accelerate particles radially, while the second term represents the 
dragging effect of the isotropic component of the radiation 
field.\footnote{This term vanishes in the case of a point-like source, 
$I(\bk)=F(r)\delta(\bk-{\bf r})$, where $F(r)$ is the radiative flux at
radius $r$.}\,
The right-hand side of equation (\ref{eq:eqT}) vanishes 
for $\Gamma=\Gamma_{\rm eq}$. Particles are accelerated away from the source 
as long as $\Gamma<\Gamma _{\rm eq}$. If $\Gamma>\Gamma_{\rm eq}$, the force 
reverses, being now directed inward.  This saturation of the Lorentz
factor is due to the aberration into the forward hemisphere of 
blueshifted photons, as seen in the electron rest-frame. A particle
will follow the $\Gamma =\Gamma_{\rm eq}$ equilibrium trajectory 
(zero-inertia limit) until, as it gets farther from the source, 
its acceleration is effectively limited by the large Doppler shift to the
red and by the $r^{-2}$ decline of the radiation flux.  

\subsection{Uniformly radiating sphere}

Consider a sphere of radius $R$ and uniform brightness $I$.  Close to the
extended source, the equilibrium Lorentz factor is 
\begin{equation}
\Gamma_{\rm eq}\sim 3^{1/4}\,{r\over R},  \label{eq:gameq}
\end{equation}
no matter how high the intensity $I$, or whether the inertia is dominated
by hadrons or leptons.  In the point-source limit ($R\rightarrow 0$) the
photons stream radially. For a steady flow, equation (\ref{eq:eqT}) reduces to
\begin{equation}
{d\Gamma \over dr}=\Gamma^2(1-\beta)^2\wiell {R\over r^2}.  \label{eq:eqps}
\end{equation}
The parameter $\wiell$ is the dimensionless compactness,
rescaled by the inertia per scattering charge,
$\wiell\equiv \ell(m_e/\mu)=L\sigma_T/(4\pi \mu c^3R)$.  It may be
expressed as $\wiell =0.5(m_p/\mu)(L/L_E)(R_S/R)$ for a source of mass $M$, 
where
$L_E=4\pi cGMm_p/\sigma_T$ is the Eddington luminosity and $R_S=2GM/c^2$
the Schwarzschild radius.  For a motion starting at radius $r$, the Lorentz
factor attained at infinity can be obtained by solving the algebraic equation 
\begin{equation}
2\Gamma_{\rm ps}^3(1+\beta^3)-3\Gamma_{\rm ps}+1=3\wiell {R\over r}, 
\end{equation}
which, in the limit $\Gamma_{\rm ps} \gg 1$, yields 
$\Gamma_{\rm ps} \sim [3\wiell R/(4r)]^{1/3}=$
$[3 L m_p R_S /(8 L_E \mu r)]^{1/3}$.
Effective acceleration of an $e^-p$ plasma requires 
a compactness that is larger by a factor $m_p/m_e$ than for
a pure pair plasma, as a conseguence of the greater inertia per unit 
cross section. 

Matching this solution with expression (\ref{eq:gameq}) yields
the asymptotic bulk Lorentz factor $\Gamma_\infty \sim \wiell^{1/4}$
(Noerdlinger 1974).  The cross-over radius beyond which the non-radial
component of the radiation field causes negligible drag is
$r_c\sim 0.8\wiell^{1/4}R$.  Larger values of $\Gamma_\infty$ can only be
obtained if the particles are injected with relativistic bulk velocities
at $r>r_c$.  Bulk motion starting with $\Gamma(r)\gg 
\Gamma_{\rm eq}(r)$ at a distance $r<r_c$ from the source is quickly
decelerated to $\Gamma \approx\Gamma_{\rm eq}$, and the excess kinetic
energy converted into a collimated beam of upscattered photons. 

\subsection{Non-uniformly radiating disk}

Infinite, Keplerian accretion disks around black holes provide an example
of non-uniform radiating sources where Compton drag is most severe. 
The surface flux distribution as a 
function of equatorial radius R is given by (Shakura and Sunyaev 1973)
\begin{equation}
F(R)={3GM{\dot M}\over 8\pi R^3}\left[1-({3R_S/R})^{1/2}\right], 
\label{eq:eqflux} 
\end{equation}
where $M$ is now the black hole mass, and $\dot M$ is the accretion rate.
The peak flux, $F_{\rm max}=3GM{\dot M}/$ $(56\pi R_{\rm max}^3)$, is reached 
at $R_{\rm max}=(49/12)R_S$. One can understand the essential features 
of the accretion disk model by approximating the flux distribution as 
\begin{equation}
F(R)= \left\{ \begin{array}{ll} F_{\rm max}R_{\rm max}^3/R^3 & 
\mbox{if $R\ge R_{\rm max}$,} \\ 0 & \mbox{if $R<R_{\rm max}$.} \end{array}
\right. \label{eq:eqapp} 
\end{equation}
In this case the equilibrium Lorentz factor increases only as 
$\Gamma_{\rm eq}\sim (r/R_{\rm max})^{1/4}$ (compared to a linear increase 
for a spherical source).
In the point-source limit, one derives $\Gamma_{\rm ps}\sim 
(0.75\wiell R_{\rm max}/r) ^{1/3}$, 
where $\wiell$ is now the nominal disk compactness, $\wiell=3GM{\dot 
M}\sigma_T/$
$(28\pi \mu c^3R_{\rm max}^2)$. Again, by matching the two limiting 
solutions, the asymptotic bulk Lorentz factor is found to scale as 
$\Gamma_\infty\sim \wiell^{1/7}$ (Kovner 1984; Phinney 1987). 
The exact solutions for a spherical uniform source and a Keplerian accretion 
disk are shown in Figure 1. The latter is a much less efficient accelerator 
than the former because of the large photon drag associated with a  
radiation source extending to infinity.

\subsection{Compression of the accelerating medium}

The radiative force acting on an external medium induces a flow that is 
strongly time-dependent.  One important effect
of acceleration to bulk Lorentz factor $\Gamma$ is the bunching of 
the accelerated material into a shell of thickness
$\Delta r \sim r/\Gamma^2$.  
Let us now solve for the density of the accelerated material. 
The equation of continuity of the hadronic component is
\begin{equation}
{d(\Gamma\tiln_p)\over dt} + 
\Gamma\tiln_p c{\partial\beta\over\partial r} = 0.
\end{equation}
Material initially at radius $r_0$ reaches a radius
\begin{equation}
r = r_0 + c\int^t_{r_0/c} \beta(t^\prime) dt^\prime = 
r_0 + c\int_0^\beta \beta^\prime 
{d\beta^\prime\over f(\beta^\prime)},
\end{equation}
at time $t$ (speed $\beta$).  Here,  $d\beta/dt = f(\beta)$.
The total duration of the acceleration is
\begin{equation}
t - {r_0\over c} = \int_0^\beta {{d\beta^\prime}\over f(\beta^\prime)}.
\end{equation}
Eliminating $r_0$ from these two equations and taking $\partial/\partial r$
at constant time $t$, one derives
$\partial\beta/\partial r = f(\beta)/c(1-\beta)$, and we have
\begin{equation}
{1\over \Gamma\tiln_p}{d\over dt}(\Gamma\tiln_p) = 
{1\over 1-\beta}{d\beta\over dt}.
\end{equation}
This integrates to 
\begin{equation}
{\Gamma \tiln_p\over n_{p\,0}} = {1\over 1-\beta} \sim 2\Gamma^2
\;\;\;\;\;\;(\Gamma\gg 1),\label{eq:newd}
\end{equation}
where $n_{p\,0}$ is the proton density in the undisturbed ambient medium.

There is a corresponding amplification of the non-radial component
of a magnetic field entrained in the flow,
\begin{equation}\label{eq:bnew}
{B\over B_0} = {\Gamma n_p'\over n_{p\,0}} = {1\over 1-\beta}.
\end{equation}

\subsection{Thin photon shell}\label{secshell}

Gamma-ray sources such as blazars and gamma-ray bursts are 
impulsive. It is instructive, therefore, to consider a photon source
that maintains a constant luminosity for a time $\Delta t \ll
r/c$.  This source can be visualized as a uniform shell of radius $r = ct$,
thickness $c\Delta t$, and total flux $F=\int F_xdx$. We focus here on the 
effects of a radiation pulse and take the photons to move radially before
scattering.

The compactness within the shell can be expressed directly in terms
of the flux of $\gamma$-rays, 
\begin{equation}
\ell\left({R\over r}\right) \rightarrow  r \sigma_T {F\over m_ec^3},
\end{equation}
and thence in terms of  a characteristic photon optical depth,\footnote{This
quantity is in fact the optical depth $\sigma_T n_\gamma(x_{\rm br}) c\Delta 
t$ multiplied by the frequency $x_{\rm br}$ at which the spectrum peaks.}
\begin{equation}
\tau_c \equiv \sigma_T \Delta t {F\over m_ec^2},\label{eq:tauc}
\end{equation}
such that $\ell(R/r) \rightarrow \tau_c (r/c\Delta t)$.
Material overtaken by the photon shell at radius $r$, and accelerated
from rest, develops a large bulk Lorentz factor when $\tau_c (m_e/\mu)\gg 1$. 
Integrating equation (\ref{eq:eqps}) in this limit yields
\begin{equation}
\Gamma^3 = {3\over 4}\wiell{R\Delta r\over r^2} =
{3\over 4}\,\tau_c\,\left({m_e\over\mu}\right)\,\left({\Delta r\over 
c\Delta t}\right)
\end{equation}
at a radius $r +\Delta r$. As the difference between
the particle speed and the speed of light decreases as $(\Delta r)^{-2/3}$, the 
accelerated matter will surf the photon shell over an extended range of radius,
$\Delta r = {2\over 3} \Gamma^2 c\Delta t$. The maximum Lorentz factor
is therefore
\begin{equation}
\Gamma_{\rm max} = {\tau_c\over 2}\,\left({m_e\over\mu}\right)
\label{eq:gams}
\end{equation}
when the shell is thin enough that the acceleration length is much less than 
the radius, ${2\over 3} \Gamma_{\rm max}^2 c\Delta t\ll r$ (see Figure 2).
Only a limited amount of material can be accelerated to this value. 
Equating the kinetic energy within a volume $(4\pi/3) r^3$ with the energy
of the photon shell implies a very high pre-existing optical depth to
Compton scattering, $\tau_T = \sigma_T \rho r/m_p \sim 6$. Since our
calculation of  $\Gamma_{\rm max}$ is valid only for small $\tau_T$, the 
ambient density at radius $r$ is bounded above by 
\begin{equation}
\rho < { \mu\over \sigma_T r} =
2\times 10^{-9}\,\left({\tau_c\over 300}\right)^{-2}\,
\left({\Delta t\over 10~{\rm s}}\right)^{-1}\,\left({\mu\over m_p}\right)^3\,
\left({3 r\over 2 \Gamma_{\rm max}^2 c\Delta t}\right)^{-1}
\;\;\;\;{\rm g~cm^{-3}}.
\label{eq:rhom}
\end{equation}
Note the strong dependence on $\mu$. We will show is Paper II that, 
even if the ambient medium
is composed of a hadronic plasma initially (as is expected in most
cosmological gamma-ray burst models), the mean mass per scattering charge
will be rapidly reduced to $\mu \sim m_e$ by pair creation.

Sidescattered photons will provide additional drag that reduces 
$\Gamma$ below the value given in equation (\ref{eq:gams}). As the photon shell
propagates ahead of an accelerating parcel of matter, it  generates
an isotropic radiation field of photon density
\begin{equation}
{n_\gamma^{\rm iso}\over n_\gamma} \sim n_s\sigma_T {r^2\over\wiell R},
\end{equation}
where $n_s\equiv n_p+2n_{e^+}$ is the density of scattering charges in the 
ambient medium at rest, and $r^2/\wiell R$ is the lengthscale over which 
isotropic scattering occurs (before the scattering medium gets accelerated to
relativistic velocities). The net effect is to introduce a second, negative 
term into equation (\ref{eq:eqps}), which becomes 
\begin{equation}
{d\Gamma\over dr} = \left({1\over 4\Gamma^2}-{4\over 3}\Gamma^2
n_s\sigma_T {r^2\over \wiell R}\right){\wiell R\over r^2}
\end{equation}
at large $\Gamma$.  Acceleration up to Lorentz factor $\Gamma_{\rm max}$
is then possible only if the ambient density is lower than
\begin{equation}
\rho <{3\tau_c\over 16\Gamma_{\rm max}^4}\,{m_e\over\sigma_T c\Delta t}.
\end{equation}
This bound is stronger by a factor $\sim \Gamma^{-1}$ than
equation (\ref{eq:rhom}): {\it in the Thomson scattering regime,
the bulk Lorentz factor is actually limited by side-scattered radiation}.
This conclusion changes when the photon source is very
hard, because side-scattered photons
are immediately consumed by pair creation (Paper II).  
A compact source of $\gamma$-rays always tends, in this sense, to maintain 
a radial photon distribution.

\subsection{Compton afterburn}\label{seccompton}

It is well known that a hot (relativistic) plasma in an anisotropic
radiation field feels a much larger radiation pressure than a cold
(non-relativistic) gas of the same inertial mass (O'Dell 1981), and tends
to drive itself away from the radiation source with momentum derived largely
from the anisotropic loss of its own internal energy. This ``Compton rocket'' 
is, however, ineffective at generating bulk relativistic motion from internal
random kinetic energy, as it is always accompanied by catastrophic Compton
cooling (Phinney 1982). Nonetheless, by integrating the momentum equation in 
the limiting case of a cold plasma we are actually underestimating its bulk 
Lorentz factor: the gas will be
continuously heated by the hard radiation field itself, hot particles
will radiate most of their energy in the direction of the radiation source,
and part of the energy of relativistic random motion will be converted into 
bulk motion. We term this supplementary source of momentum the 
`Compton afterburn'.
The afterburn can be largely neglected in the case of pure KN scattering
without pair creation (\S\,4.1).
(In Paper II we will consider a situation in which relativistic $e^+e^-$
pairs are injected into an anisotropic photon beam by collisions between 
side-scattered photons and the main beam -- thereby inducing a net force on 
the plasma as they Compton cool.) Nonetheless, we discuss it here for 
completeness.
    
The afterburn is strongest when the radiative force on the relativistic,
cooling particles  is coupled to the rest of the fluid through a background
magnetic field.  Above a critical flux density which we estimate, this
allows the colder particles to dominate the inertia of the flow, and 
the momentum gained by Compton scattering to be more effectively absorbed 
rather than being radiated away. The afterburn is easiest to analyze when the
period of a cyclotron orbit is much shorter than the Compton cooling time.
Then the linear momentum imparted to one charge is effectively shared with
the others before the charge loses its excess energy in the bulk frame.
We also assume that the acceleration time is long compared with the
Compton cooling time, so that the proportion of energetic particles
is small.  This is the case when the energetic charges are continually
regenerated, e.g., by pair creation.  Finally, we assume that the magnetic
field is non-radial (as is appropriate to the rest frame of
a plasma that has been accelerated to relativistic speed).  We divide
the velocity of the charge into components $\beta_{e\perp}$ and
$\beta_{e\,\parallel}$ perpendicular and parallel to ${\bf B}$, and denote
by $\eta$ the angle between the cyclotron motion and the radial direction.

The charge is immersed in a collimated photon beam of energy flux 
\begin{equation}
F_\Gamma = {1\over 2\Gamma^2}F={1\over 2\Gamma^2}\int F_xdx\;\;\;\;\;\;(\Gamma 
\gg 1)\label{eq:fgam}
\end{equation}
in the boosted frame of the bulk flow.  It feels a radial force (O'Dell 1981)
\begin{equation}
{d\over dt}\left({p_r\over m_ec}\right) = {\sigma_T F_\Gamma\over m_ec^2}
(1-\beta_{e\perp}\cos\eta)\Bigl[\gamma_e^2(1-\beta_{e\perp}\cos\eta)
(-\beta_{e\perp}\cos\eta) + 1\Bigr].
\end{equation}
The net radial momentum imparted to the cooling charge is obtained by
averaging over a cyclotron orbit, $\eta \rightarrow \eta + 2\pi$,
\begin{equation}
\biggl\langle{d\over dt}\left({p_r\over m_ec}\right)\biggr\rangle = 
{\sigma_T F_\Gamma\over m_ec^2}\Bigl(\gamma_e^2\beta_{e\perp}^2+1\Bigr).
\end{equation}
At the same time, the net energy lost to Compton cooling is
\begin{equation}
\biggl\langle{d\gamma_e\over dt}\biggr\rangle 
= -{\sigma_T F_\Gamma\over m_ec^2}\gamma_e^2
\left({1\over 2}\beta_{e\perp}^2 + \beta_e^2\right).
\end{equation}

When the energetic charges are created with $\beta_{e\,\parallel} = 0$
(e.g., pair creation between a hard radial photon and a soft side-scattered
photon), the cooling rate is $d\gamma_e/dt = -(3\sigma_T F_\Gamma/2m_ec^2)\,
\gamma_e^2\beta_e^2$  and the radial momentum gained as the particle cools is
\begin{equation}
{d(p_r/m_ec)\over d\gamma_e} = -{2\gamma_e^2\over 3(\gamma_e^2-1)}.
\end{equation}
To leading order in $\gamma_e$,
\begin{equation}
{p_r\over m_ec} \simeq {2\over 3}\gamma_e.\label{eq:prad}
\end{equation}
In this case, the source of energy for the charge is external to 
the accelerating medium.  The charge carries an excess lab-frame momentum
$(2\Gamma)\gamma_e m_ec$, which is supplemented by (\ref{eq:prad}) 
to yield a total ${5\over 3}(2\Gamma)\gamma_e m_ec$ (in the limit
$\gamma_e \gg 1$).  

These expressions change slightly when the energetic charges are
distributed isotropically, and reduce to those found by O'Dell (1981).
The rate of Compton cooling is $d\gamma_e/dt$ $=-(4\sigma_T F_\Gamma/3m_ec^2)$
$\,\gamma_e^2\beta_e^2$, as usual, and
\begin{equation}
{d(p_r/m_ec)\over d\gamma_e} = -{\gamma_e^2+{1\over 2}\over 2(\gamma_e^2-1)}.
\end{equation}
The {\it net} radial momentum acquired is 
\begin{equation}
{p_r\over m_ec} \simeq {1\over 2}\gamma_e,\label{eq:pradb}
\end{equation}
only $3/10$ of the value obtained for charges that are created
moving parallel to the photon beam.  

These calculations assume that a high energy particle will execute 
many cyclotron orbits before Compton cooling.  Let us check when
this assumption is valid.  The rest frame magnetic flux density is related to
the initial (non-radial) field before acceleration by
\begin{equation}
B' = {B_0\over \Gamma(1-\beta)} \simeq 2\Gamma\,B_0
\end{equation}
(eq. \ref{eq:bnew}).  Each fresh (radially-moving) particle has a
rest-frame energy $\gamma_e m_ec^2 \simeq E/2\Gamma$, and so it takes a time
\begin{equation}
t'_{\rm cyc} = {\gamma_em_ec\over eB'} \simeq 
{1\over 4\Gamma^2}\,{E\over eB_0c}
\end{equation}
to complete one radian of a Larmor orbit.  The Compton cooling time is
\begin{equation}
t'_{\rm cool} = {2m_ec^2\over 3\gamma_e\sigma_T F_\Gamma}.
\end{equation}
The energy flux in a thin photon shell is related to the characteristic
optical depth $\tau_c$ (eq. \ref{eq:tauc}) through $\sigma_T F/m_ec^2
= \tau_c/\Delta t$, and so the ratio of these two timescales is
\begin{equation}
{t'_{\rm cyc}\over t'_{\rm cool}} \sim {3\over 32 \Gamma^5} \tau_c\,
\left({E\over m_ec^2}\right)^2\,\left({m_ec\over eB_0\Delta t}\right)
= 2\times 10^{-4}\,{\tau_c\over\Gamma^5}\,\left({E\over m_ec^2}\right)^2\,
\left({B_0\over 3\times 10^{-6}~{\rm G}}\right)^{-1}\,
\left({\Delta t\over 10~{\rm s}}\right)^{-1}.\label{eq:tcomp}
\end{equation}

The compression of the accelerating medium has a related consequence:
the mean energy $\gamav$ of the scattering charges in the bulk frame
increases due to adiabatic heating, at a rate
$\gamav^{-1}(d\gamav/dt) = {1\over 3}\tiln^{-1}(d\tiln/dt)
= {1\over 3}\Gamma^{-1}(d\Gamma/dt)$ in the lab frame.  A balance between
compressional heating and Compton cooling results in an equilibrium energy
$\gamav$ that is only mildly relativistic.  As a result, we will assume that
the heating process has isotropized the momenta of the pairs in the bulk frame:
\begin{equation}
{1\over\gamav}{d\gamav\over dt} = {1\over 3\Gamma}{d\Gamma\over dt}
- {4\over 3}{\gamav\over t_{\rm cool}^0}.
\end{equation}
The reference (lab-frame) cooling time is $t_{\rm cool}^0 = 
m_ec^2/\sigma_T \Gamma F_\Gamma \simeq (\Gamma x_{\rm br}/2\tau_c)\Delta t$.
The acceleration rate is decreased by a factor $\sim \gamav^{-1}$ from
the value calculated above for a cold plasma: 
\begin{equation}
t_{\rm accel} = {\Gamma\over d\Gamma/dt} \simeq \gamav\,t_{\rm accel}^0.
\end{equation}
>From  equation (\ref{eq:eqps}), we have
\begin{equation}
{t_{\rm cool}^0\over t_{\rm accel}^0} = {1\over 2\Gamma^2}{m_e\over\mu}
\end{equation}
at $\Gamma \gg 1$.
Even in the case of a pair-loaded plasma ($\mu \simeq m_e$), one
sees that energetic particles cool faster than the bulk flow accelerates.
The same conclusion holds for scattering in the Klein-Nishina regime
(\S\,4), but not when pair creation drives the acceleration (Paper II).

\section{Klein-Nishina regime:  monoenergetic spectrum}

When $x'\gta 1$ recoil can no longer be neglected. In the general case, the 
integral over the scattering angle $\chi'$ of the Compton force on 
particles moving along the {\bf r}-axis,  
\begin{equation}
{d\over dt}\langle \Gamma\beta\rangle
=- {1\over \mu c^2} \int {x_s \cos \theta_s-x \cos 
\theta \over x}\,(1-\beta \cos \theta)I_xdxd\sigma d\Omega, \label{eq:comp}
\end{equation} 
can be performed analitically by making the variable change $\chi'\rightarrow 
x_s'$, and integrating the resulting polynomial in $x_s'$. The mean rate of 
momentum transfer can then be written as
\begin{equation}
{d\over dt}\langle\Gamma\beta\rangle
={\sigma_T\over \mu c^2} \int (1-\beta \cos\theta)
K(x') \left[\cos\theta +{\Gamma(\cos\theta-\beta)\over x(1-\beta\cos\theta)}
\right]I_xdxd\Omega, \label{eq:eqpkn}
\end{equation}
where $K(x')\equiv \int x_s'(1-\cos\chi')d\sigma/\sigma_T$ can be 
evaluated in closed form, 
\begin{equation}
K(x')={3\over 4 x'^2}\left[{x'^2 -2x'-3\over 2x'}\ln(1+2x')
+{-10x'^4 +51x'^3 +93x'^2 +51x'+9\over 3(1+2x')^3}\right] \label{eq:int}
\end{equation}
(cf. Blumenthal 1974). This function has limiting behavior 
\begin{equation}
K(x')\approx {3\over 8 x'}[\ln(2x')-5/6+....]
\end{equation}
for $x'\gg 1$, and
\begin{equation}
K(x')\approx x'(1-21x'/5+.....)
\end{equation}
for $x'\ll 1$.  The first term in the last expansion, when substituted back into
equation (\ref{eq:eqpkn}) in the limit $x'\ll 1$, reproduces the standard 
Thomson results of equation (\ref{eq:eqT}) (for $\mu=m_e$). 
\footnote{The force on a $e^-p$ plasma at rest relative to a 
point-source of mass $M$
and monoenergetic flux $F(r)$ is, from equation (\ref{eq:eqpkn}), 
$d\langle\Gamma\beta\rangle/dt =(\sigma_T/m_pc^2) F(r) K(x) (1+x)/x$. The
Eddington luminosity for which this force just balances the gravitational
attraction at all radii $r$ is then given by 
$$
L_E={4\pi c G M m_p\over \sigma_T} \left[{x\over K(x) (1+x)}\right],
$$
where the term in parenthesis (equal to 2.4, 5.9, and 11.8 for $x=0.5, 3,$ 
and 10, respectively) is the relativistic correction factor to the classical 
expression in the Thomson limit (Blumenthal 1974).}\, 

Figure 3 shows the integrand function in equation (\ref{eq:eqpkn}), $P
\equiv (1-\beta \cos\theta) K(x') \{\cos\theta +\Gamma(\cos\theta-\beta)/
[x(1-\beta\cos\theta)]\}$, versus incident photon angle for different photon
energies and particle valocities. The aberration effect of outward-directed 
photons into the forward hemisphere (as observed in the electron rest-frame)
is clearly seen, together with the reduced radiative force as $\gamma$-ray 
photons are preferentially scattered in the forward 
direction. In the Thomson limit, photons at angles $\cos \theta<\beta$ with 
respect to the direction of motion are seen in the rest-frame of the 
electron as blueshifted and inward directed, and work to decelerate the 
flow ($P<0$). The same criterion does not apply in Klein-Nishina, as 
blueshifted, aberrated photons experience a decreased relativistic
cross section and are scattered 
less efficiently. When $\beta=0.8$, for example, the rate of momentum 
transfer becomes negative at angles $\cos\theta<0.71, 0.35,$ and
0.13 for photons with energy $x=0.5, 3,$  and 10, respectively.   

The complexity of the KN cross section foils analytic calculations
of the particle trajectories. We have numerically integrated equation 
(\ref{eq:eqpkn}) with the Runge-Kutta method assuming a monoenergetic photon
spectrum. The equilibrium Lorentz factor as a function of distance from a 
radiating sphere and a disk is shown in Figure 4 for different energies of the 
incoming photons. While in the Thomson limit the radiation drag imposes a 
linear relation between $\Gamma_{\rm eq}$ and distance $r$ from a spherical
source,
in Klein-Nishina the importance of this braking force 
is reduced: the zero-inertia limit is reached then at higher particle 
velocities. The net momentum transfer in each interaction with a $\gamma$-ray
photon is significantly lower, however, and it is actually more difficult 
(i.e. larger compactnesses are needed) for an electron to approach the 
zero-inertia limit (Figure 4). 
In Figure 5 the asymptotic Lorentz factor at infinity is
depicted as a function of source compactness for different incoming photon 
energies. A comparison with the Thomson limit 
shows that $\gamma$-ray photons with $x=10$ (say) are significantly 
more efficient at accelerating particles only for compactness $\wiell\gta 200$, 
and more so in the extended disk case (i.e. when the isotropic radiation 
field is larger) than for a spherical source. In the general case, the detailed 
dynamics of an the plasma will depend on the energy spectrum of the incoming 
radiation (Paper II). 

The KN reduction of the photon drag can be easily quantified in the extreme 
case of a jet moving with relativistic speed towards a point-source 
of $\gamma$-rays. From equation (\ref{eq:eqpkn}) we derive
\begin{equation}
{d\over dt}\langle\Gamma\beta\rangle=
-{\sigma_T F(r)\over \mu c^2}\, (1+\beta^2)\Gamma^2
\left[{K(x')\over \Gamma^2 (1+\beta)}\left(1+{\Gamma\over x}\right)\right],
\end{equation}
where $x'=x\Gamma(1+\beta)$.
The term in parenthesis, equal to 0.094, 0.016, and 0.006 for $\Gamma=2$ and 
$x=0.5, 3,$ and 10, respectively, is the relativistic correction to the 
braking force in scattering Thomson. 

\subsection{Compton heating}

Together with momentum, energy will be transferred from the radiation
field to the gas at a rate
$$
\biggl\langle{dE\over dt}\biggr\rangle =-{1\over \mu c^2} \int {x_s-x\over x}\,
(1-\beta \cos \theta) I_xdxd\sigma d\Omega
$$ 
\begin{equation}
~~~~~~~~~~~~~~~~~~~~~~~~~~~={\sigma_T\over \mu c^2} \int (1-\beta \cos\theta)
K(x') \left[1+{\Gamma\beta(\cos\theta-\beta)\over x(1-\beta \cos\theta)}
\right]I_xdxd\Omega. 
\label{eq:Ecomp}
\end{equation} 
Equations (\ref{eq:eqpkn}) and (\ref{eq:Ecomp}) assume that all the particles 
in a small volume element of gas (that moves with bulk Lorentz factor $\Gamma$) 
are at rest with respect to that volume 
element. Because of recoil in each collision, however, the 
plasma will be Compton heated (hence $\langle dE/dt \rangle \neq d\Gamma/dt$ in 
eq. \ref{eq:Ecomp}) 
to relativistic temperatures, and in general these equations should be 
integrated over the appropriate distribution function. 
\footnote{In the absence of any other cooling or heating
mechanisms, a thermal gas will be driven towards the Compton 
temperature $T_C=m_ec^2\langle x\rangle/(4k)$. For a power-law spectrum with 
energy indices $\alpha=0$ and $-1.5$ below and above the break frequency 
$x_{\rm br}=1$
(characteristic of gamma-ray bursts), one derives $\langle x\rangle=0.74$
and $T_C=1.1\times 10^9\,$K.}\, The integral can 
can solved numerically for a momentum distribution that is isotropic in the 
electron rest-frame (e.g. a relativistic Maxwellian) provided the relaxation 
time of the plasma is short compared with the dynamical time. 

In the case of a cold plasma at rest in an {\it 
isotropic} photon bath, the Compton heating rate per unit volume can be 
written, from equation (\ref{eq:Ecomp}) with $x'=x$, 
\begin{equation}
H=4\pi n_s\sigma_T\int \left[K(x)\over x\right] xI_xdx,
\label{eq:Hcomp}
\end{equation} 
where the term $K(x)/x$ (equal to 0.28, 0.04, and 0.008 for $x=0.5, 3,$
and 10, respectively) is the KN reduction factor to the classical Thomson 
formula. It has been recently pointed out that Compton 
heating of electrons by hard X-ray background photons may provide a 
significant energy source for the intergalactic medium (Madau \& Efstathiou 
1999). It is crucial in the cosmological context to use equation 
(\ref{eq:Hcomp}) rather the Thomson limit, as relativistic corrections become
increasingly more important at early times when the peak in 
spectral power of the X-ray background (observed today at $\sim 30\,$ keV) 
is blueshifted to higher energies.  

It is also worth noting that the efficiency of the Compton afterburn should be
further diminished in the KN regime, as the decreased relativistic cross section
reduces the anisotropy (between highly scattering electrons moving towards
the source  and those, less scattering, moving away) that ultimately drives
the effect (Phinney 1982; see also Renaud \& Henry 1998). The afterburn 
may be a supplementary source of momentum if the energetic particles are 
injected by an external energy source, and if they execute many Larmor orbits 
while cooling (\S \ref{seccompton}). However, in the present context 
(in which the
particles are assumed cold before scattering) only a small increase in
the bulk momentum will result from the afterburn effect.  Even if we suppose
that {\it all} the energy transferred from the radiation field to
the gas is actually converted into bulk motion, and integrate equation
(\ref{eq:Ecomp}) with
$\langle dE/dt \rangle=d\Gamma/dt$, we get asymptotic Lorentz factors that are
at most ten percent higher than those computed from the momentum equation.

\acknowledgments
We have benefited from stimulating discussions with G. Blumenthal,  
G. Ghisellini, and M. Rees. Support for this work was provided by NSF through 
grant 
PHY94-07194 (P. M.), and by the Sloan foundation (C. T.).

\references

Abramowicz, M.~A., \& Piran, T. 1980, \apj, 241, L7

Bloemen, H., \etal 1995, A\&A, 293, L1

Blumenthal, G. R. 1974, \apj, 188, 121

Fichtel, C.~E., \etal 1994, ApJS, 94, 551

Kovner, I. 1984, A\&A, 141, 341

Luo, Q., \& Protheroe, R. J. 1999, MNRAS, in press (astro-ph/9901223)

Madau, P., \& Efstathiou, G. 1999, ApJ, 517, L9 

Mandrou, P., \etal 1994, ApJS, 92, 343

Melia, F., \& K\"{o}nigl, A. 1989, \apj, 340, 162

Noerdlinger, P.~D. 1974, \apj, 192, 529

O'Dell, S.~L. 1981, \apj, 243, L147

Phinney, E.~S. 1982, \mnras, 198, 1109

Phinney, E.~S. 1987, in Superluminal Radio Sources, ed. J.~A.
Zensus \& T.~J. Pearson (Cambridge: Cambridge Univ. Press), 301

Renaud, N., \& Henri, G. 1998, \mnras, 300, 1047

Rybicki, G.~B., \& Lightman, A.~P. 1979, Radiative Processes in Astrophysics 
(New York: Wiley)

Shakura, N.~I., \& Sunyaev, R.~A. 1973, A\&A, 24, 337

Sikora, M., Sol, H., Begelman, M.~C., \& Madejski, G.~M. 1996, \mnras, 280, 781 

Sikora, M., \& Wilson, D. 1981, \mnras, 197, 529

Svensson, R. 1982, \apj, 258, 335 

Thompson, C., \& Madau, P. 1999, submitted to the ApJ (Paper II) 

Ulmer, M.~P. 1994, ApJS, 90, 789

\vfill\eject

\begin{figure}
\plotone{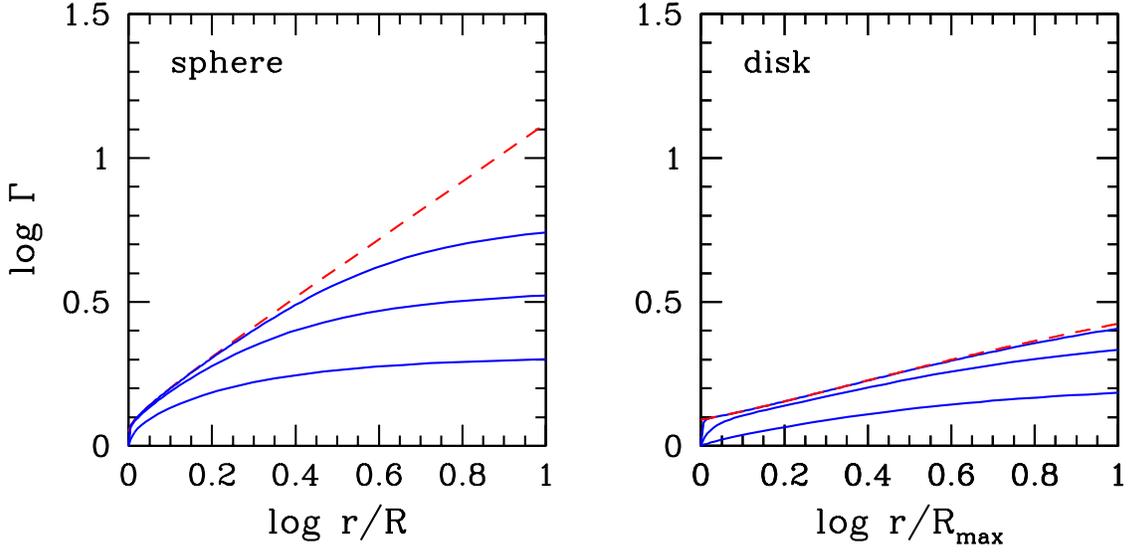}
\caption{{\it Left:} Bulk Lorentz factor of a test particle  
as a function of distance $r$
from a uniformly radiating sphere of radius $R$ and compactness $\wiell$. The 
equation of motion has been integrated in the Thomson regime assuming the 
particle is initially at rest. From top to bottom: $\wiell=1000, 100, 10$ 
({\it solid lines}). {\it Dashed line}: Zero-inertia limit. 
{\it Right:} Same but for an infinite, Keplerian accretion disk (see text
for details). 
\label{fig1}}
\end{figure}

\begin{figure}
\plotone{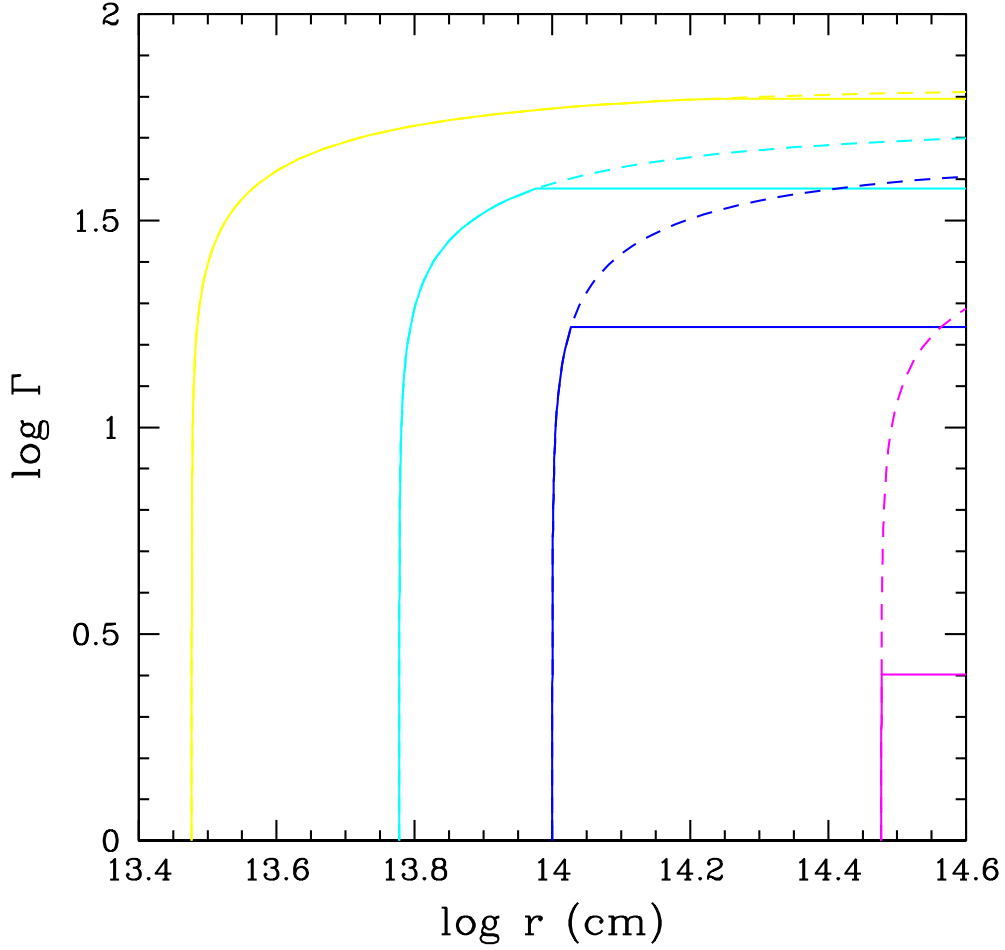}
\caption{Bulk Lorentz factor of a test particle ($\mu=m_p$)  
as a function of distance $r$ (in cm) from a point-like source of luminosity 
$10^{52}$ ergs s$^{-1}$. The 
equation of motion has been integrated in the Thomson regime assuming the 
particle to be initially at rest. {\it Solid curves}: radiation pulse of 
duration 1 sec. The radiative force vanishes when the photon shell moves past 
the particle.
{\it Dashed curves}: steady source.
\label{fig2}}
\end{figure}

\begin{figure}
\plotone{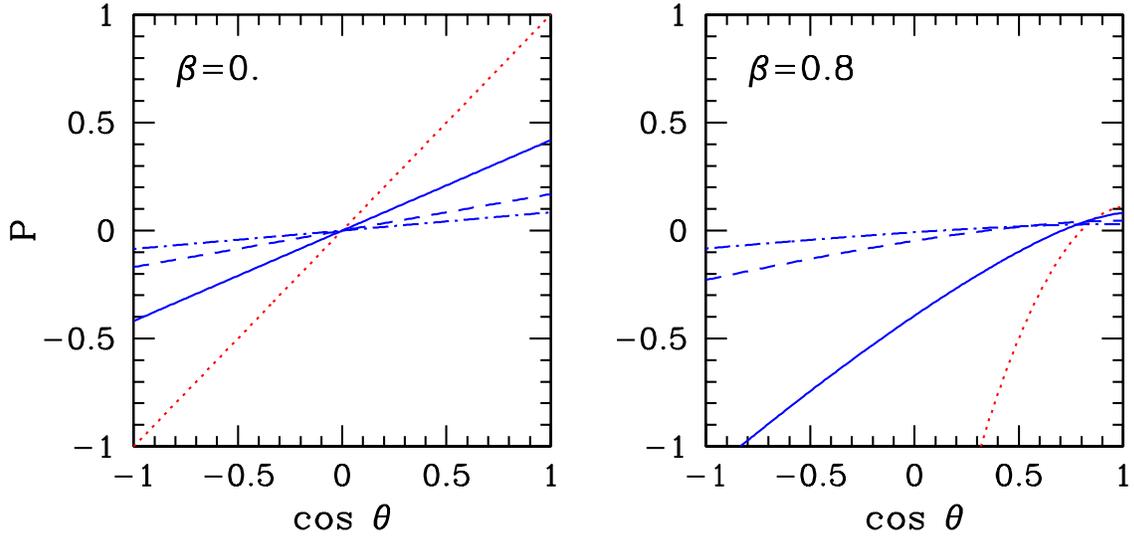}
\caption{Specific rate of momentum transfer per unit solid angle 
(see eq. \ref{eq:eqpkn}), $P
\equiv (1-\beta \cos\theta) K(x') \{\cos\theta+\Gamma(\cos\theta-\beta)/
[x(1-\beta\cos\theta)]\}$, versus incident photon angle for different photon
energies $x$ (in units of $m_ec^2$) and particle valocities $\beta$ (in units
of $c$). {\it Solid line:} $x=0.5$. {\it Dashed line:} $x=3$. 
{\it Dash-dotted line:} $x=10$. {\it Dotted line:} Thomson limit. 
\label{fig3}}
\end{figure}

\begin{figure}
\plotone{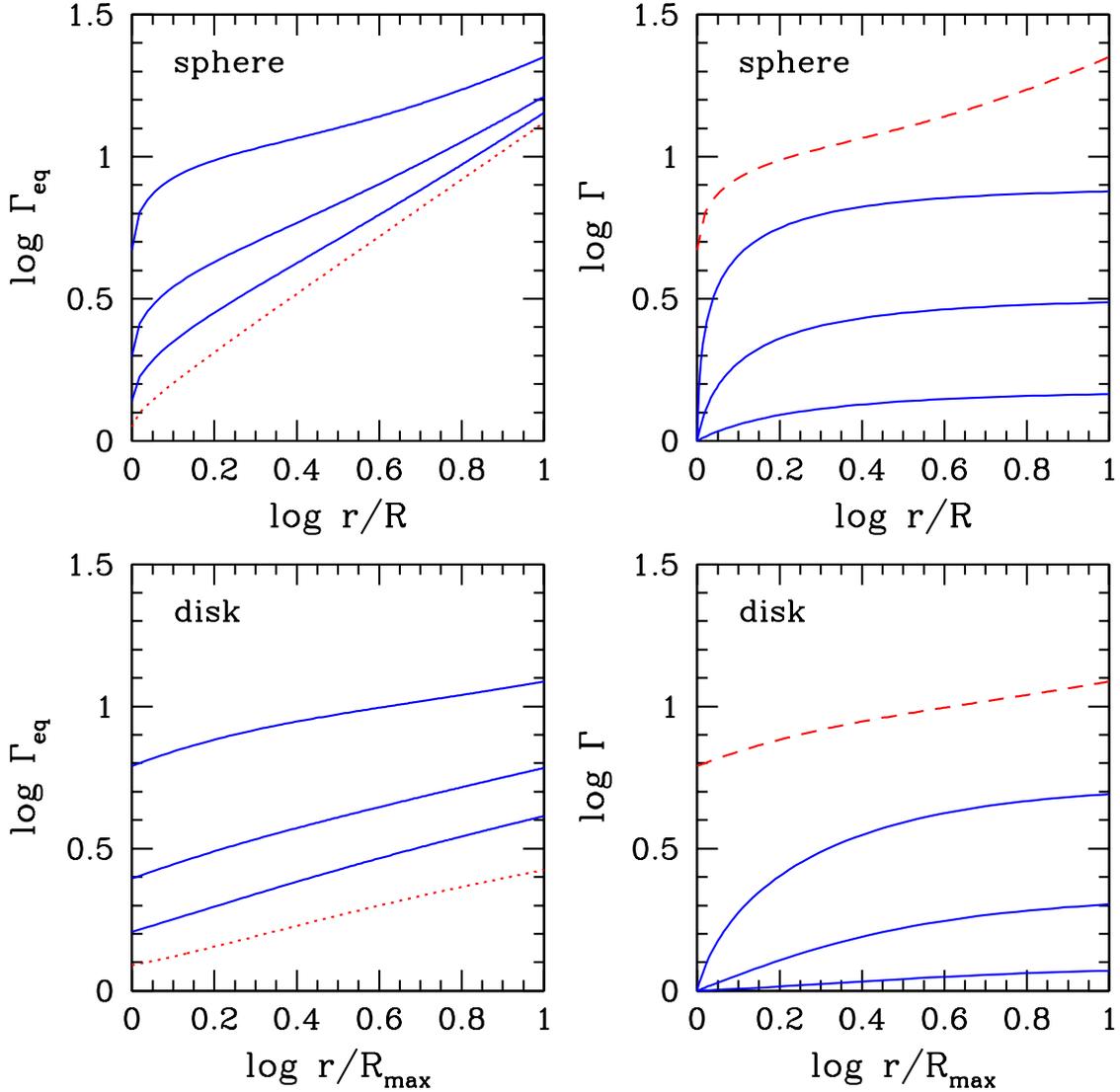}
\caption{{\it Top left:} Equilibrium 
Lorentz factor (zero-inertia limit) of a test 
particle as a function of distance $r$ from a uniformly radiating sphere 
of radius $R$. The photon spectrum is monoenergetic and the equation of motion
has been integrated in the Klein-Nishina regime. {\it Solid lines:} Values 
obtained for different energies of the incoming photons: $x=10, 3, 1$ 
(from top to bottom). {\it Dotted line:} Thomson limit. {\it Top right:} Bulk 
Lorentz factor as a function of distance. The equation of motion has been 
integrated using the relativistic cross section and assuming 
the particle to be initially at rest. The radiation spectrum is monoenergetic 
with photon energy $x=10$. From top to bottom: 
$\wiell=1000, 100, 10$ ({\it solid lines}). {\it Dashed line}: Zero-inertia 
limit. Note how, in the relativistic limit, the particle velocity saturates 
to its asymptotic value at infinity much closer to the radiation source than
in the Thomson regime (cf. Fig. 1). {\it Bottom:} Same for a Keplerian
accretion disk.
\label{fig4}}
\end{figure}

\begin{figure}
\plotone{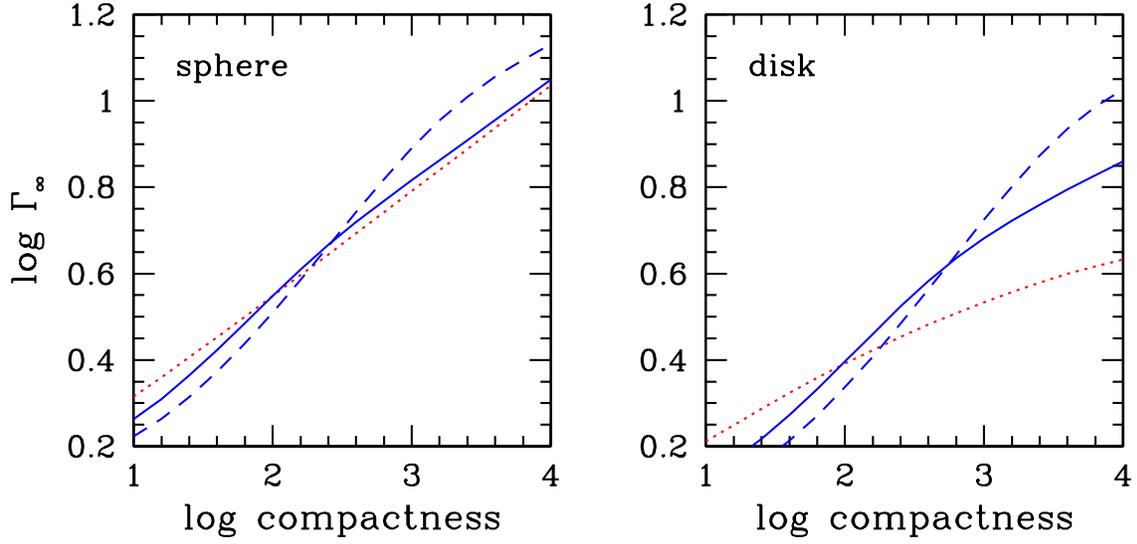}
\caption{Asymptotic Lorentz factor at infinity as a function of compactness 
for a particle initially at rest. {\it Solid line:} KN cross section, 
incoming photon energy
$x=3$. {\it Dashed line:} Same for $x=10$. {\it Dotted line:} Thomson limit. 
{\it Left:} Uniformly radiating sphere. {\it Right:} Keplerian accretion disk.  
\label{fig5}}
\end{figure}

\end{document}